\documentclass{PoS}
\usepackage{mathrsfs}
\usepackage{lipsum}
\usepackage{url}

\title{Semileptonic $B$ to $D$ decays at nonzero recoil with 2+1 flavors of
improved staggered quarks}

\ShortTitle{Semileptonic $B\rightarrow Dl\nu$ decays at nonzero
recoil}

\author{\speaker{Si-Wei Qiu}\\
        Department of Physics and Astronomy, University of Utah, Salt Lake City, UT84112, USA\\
        E-mail: \email{siwei.qiu@utah.edu}}
\author{Carleton DeTar\\
        Department of Physics and Astronomy, University of Utah, Salt Lake City, UT84112, USA\\
        E-mail: \email{detar@physics.utah.edu}}
\author{Daping Du\\
        Department of Physics and Astronomy, University of Iowa, Iowa City, IA 52240, USA\\
        E-mail: \email{ddu@illinois.edu}}
\author{Andreas S. Kronfeld\\
        Theoretical Physics Department, Fermi National Accelerator Laboratory, Batavia, IL 60510, USA\\
        E-mail: \email{ask@fnal.gov}}
\author{Jack Laiho\\
        SUPA, Department of Physics and Astronomy, University of Glasgow, Glasgow, Scotland, UK\\
        E-mail: \email{jlaiho@fnal.gov}}
\author{Ruth S. Van de Water\\
        Physics Department, Brookhaven National Laboratory, Upton, New York 11973, USA\\
        E-mail: \email{ruthv@bnl.gov}}
\author{(Fermilab Lattice and MILC Collaborations)}

\abstract{The Fermilab Lattice-MILC collaboration is completing a
comprehensive program of heavy-light physics on the MILC
(2+1)-flavor asqtad ensembles with lattice spacings as small as
0.045 fm and light-to-strange-quark mass ratios as low as 1/20.  We
use the Fermilab interpretation of the clover action for heavy
valence quarks and the asqtad action for light valence quarks. The
central goal of the program is to provide ever more exacting tests
of the unitarity of the CKM matrix. We give a progress report on one
part of the program, namely the analysis of the semileptonic decay
$B$ to $D$ at both zero and nonzero recoil. Although final results
are not presented, we discuss improvements in the analysis methods,
the statistical errors, and the parameter coverage that we expect
will lead to a significant reduction in the final error for
$|V_{cb}|$ from this decay channel. }

\FullConference{ The XXIX International Symposium on Lattice Field Theory - Lattice 2011\\
July 10-16, 2011\\
Squaw Valley, Lake Tahoe, California}

\begin{document}

\section{ Introduction\label{sec:In}}
Precision tests of the standard model from flavor factory and
intensity frontier experiments can reveal new physics even if the
accelerator energy isn't sufficient to create the new particles
associated with the new physics. Here preliminary results of a
lattice-QCD calculation of the nonzero-recoil form factor for the
semileptonic process $B \rightarrow D\ell\overline{\nu}$ are
presented. The principal goal is to determine $|V_{cb}|$ to a high
precision. The theoretical uncertainty on $|V_{cb}|$ limits the
precision of the unitarity triangle constraint from neutral kaon
mixing.

There are several methods for determining $|V_{cb}|$.  (1) From the
inclusive decay $b\rightarrow c \ell \overline{\nu}$, perturbation
theory and the operator product expansion provide an estimate of the
inclusive decay rate, which is then combined with the measured decay
rate to obtain $|V_{cb}|$ \cite{Boyd:1996hy}.  (2) From the
exclusive semileptonic decay $B\rightarrow D^* \ell \overline{\nu}$,
lattice-QCD methods provide the hadronic contribution to the decay
rate \cite{Shoji}. For the special case of zero recoil, we have
presented an unquenched calculation for this process three years ago
~\cite{Bernard:2008dn} with a preliminary update reported at the
Lattice 2010 ~\cite{Bailey:2010gb} and CKM 2010 conferences
\cite{Mackenzie} (3) Likewise, from the exclusive semileptonic decay
$B\rightarrow D\ell \overline{\nu} $, lattice-QCD methods provide
the hadronic contribution \cite{Divitiis}. Here we report the first
unquenched lattice-QCD calculation for this process at nonzero
recoil.

\begin{figure}[b]
\centering
\includegraphics[width=0.5\textwidth]{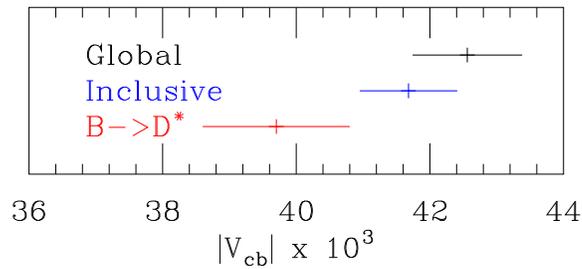}
\caption{\label{fig:fi2} The central values and errors
resulting from three determinations of $|V_{cb}|$:
(1) A ``global fit'' to the unitarity triangle including all inputs
except $|V_{cb}|$ \cite{global} \cite{web}, (2) inclusive measurements
summarized by the HFAG \cite{web1}, and (3) our $B\rightarrow D^*$
lattice-QCD calculation \cite{Bailey:2010gb,Mackenzie}.}
\end{figure}

At present there is a 1.6 $\sigma$ disagreement between the value of
$|V_{cb}|$ determined from inclusive decays and our recent
preliminary value based on the decay $B\rightarrow D^* \ell
\overline{\nu}$ at zero recoil, as illustrated in
Fig.~\ref{fig:fi2}.  Thus, further cross checks of and improvements
in the theoretical calculations are needed. The value of $|V_{cb}|$
can be determined from the ratio of the measured decay rate to the
calculated hadronic form factor at any chosen recoil energy.
Typically, the uncertainty in the experimental measurement increases
as the recoil energy vanishes \cite{Bartelt}, whereas the
uncertainty in the lattice-QCD calculation decreases
\cite{Kronfeld:2000ck,Ar,Divitiis}. Thus, to minimize the
uncertainty in the ratio, it is best to combine results of the
experimental measurement with those of the lattice-QCD calculation over
the full range of available recoil energies.  For this reason we
undertake an analysis over a broad range.

\section{\label{sec:VB} Methodology for determining $|V_{cb}|$}
The hadronic weak matrix element for this process is commonly parameterized as
\begin{eqnarray}
 \langle D(p^{\prime})|\mathcal {V}_{\mu}|B(p)\rangle=\sqrt{m_Bm_D}[h_+^{B\rightarrow D}(w)
 (v+v^{\prime})_{\mu}+h_-^{B\rightarrow D}(w)
 (v-v^{\prime})_{\mu}]\ .
\end{eqnarray}
Here, $w$ is velocity transfer, $w=v\cdot v^{\prime}$. The
differential rate for the semileptonic decay $B\rightarrow D\ell
\overline{\nu}$ is

\begin{eqnarray}
\frac{d\Gamma(B\rightarrow
Dl\overline{\nu})}{dw}=\frac{G_F^2|V_{cb}|^2}{48\pi^3}(m_B+m_D)^2m_D^3[w^2-1]^{3/2}|\mathcal{G}_{B\rightarrow
D}(w)|^2 \ ,
\end{eqnarray}
where $\mathcal {G}_{B\rightarrow D}(w)$ is defined as
\begin{eqnarray}
\mathcal{G}_{B\rightarrow D}(w)=h^{B\rightarrow
D}_+(w)-\frac{m_B-m_D}{m_B+m_D}h^{B\rightarrow D}_-(w) \ .
\end{eqnarray}
Thus, once we know $h^{B\rightarrow D}_+$ and $h^{B\rightarrow
D}_-$, we can determine the form factor $\mathcal{G}_{B\rightarrow
D}(w)$. Then, combining it with the result of the decay rate from
experiment, we will obtain $|V_{cb}|$. In order to construct
$h^{B\rightarrow D}_+$ and $h^{B\rightarrow D}_-$, we need the
following quantities:
\begin{eqnarray}
h_+(w)&=&R_+({\bf p})[1-x_f({\bf p}) R_-({\bf p})], \\
h_-(w)&=&R_+({\bf p})[1-R_-({\bf p})/x_f({\bf p})], \\
R_+({\bf p})&=&\langle D({\bf p})|\mathcal{V}^4|B(0)\rangle, \label{threepts1}\\
R_-({\bf p})&=&\frac{\langle D({\bf p})|\mathcal{V}^1|B(0)\rangle}
   {\langle D({\bf p})|\mathcal{V}^4|B(0)\rangle}, \\
x_f({\bf p})&=&\frac{\langle D({\bf p})|\mathcal{V}^1|D(0)\rangle}%
{\langle D({\bf p})|\mathcal{V}^4|D(0)\rangle}\ ,
\label{threepts}
\end{eqnarray}
where $\mathcal{V}^\mu$ is the continuum hadronic weak vector current.

In our calculation we use the local (nonconserved) lattice vector
current $V_\mu(x) = \bar\Psi_b(x) i\gamma_\mu \Psi_c(x)$ with ${\mathcal
  O}(a)$ improved heavy quark fields $\Psi_h(x)$ following
\cite{ElKhadra:1996mp}, and we renormalize it following a partly nonperturbative method,
namely
\begin{equation}
{Z_{Vbc}} = \rho_V\sqrt{Z_{Vbb}Z_{Vcc}}\ ,
\end{equation}
where the flavor-diagonal renormalization coefficients are computed
nonperturbatively on the lattice via the conditions
\begin{eqnarray}
Z_{Vbb}\langle B | V^4 | B\rangle&=&1,\\
Z_{Vcc}\langle D | V^4 | D\rangle&=&1,
\end{eqnarray}
and $\rho_V$ is computed perturbatively \cite{Ar}.  Because of
cancellations among similar loop diagrams, we expect that $\rho_V$
is nearly equal to 1 \cite{Ar}. Preliminary results presented here
omit the $\rho_V$ factor.

At zero recoil ($w = 1$) we use the double ratio method
\cite{Shoji}:
\begin{eqnarray}
|h_+(1)|^2=\rho_V^2\frac{\langle D|V^4|B\rangle\langle
B|V^4|D\rangle}{\langle D|V^4|D\rangle\langle B|V^4|B\rangle} \ \ .
\end{eqnarray}
This ratio suppresses a large part of the statistical fluctuations,
and it builds in the current renormalization
$\sqrt{Z_{V^{bb}}Z_{V^{cc}}}$.

\section{\label{sec:Lat} Results}

We report on results from an analysis of a large set of gauge field
ensembles generated in the presence of $2+1$ flavors of improved
staggered (asqtad) quarks \cite{Bernard:1999xx,Bazavov:2009bb}.
Further ensembles will be included in the future.  Some key ensemble
parameters are listed in Table ~\ref{table1}.
\begin{table}

\label{table1}
\begin{minipage}{0.48\linewidth}
\centering
\begin{tabular}{lll}
\hline
$a$(fm) & size& $m_l/m_h$\\
\hline
$\approx$0.15 & $16^3\times 48$ & 0.2 \\
$\approx${\bf 0.12} & $20^3\times 64$ &  0.14 \\
$\approx${\bf 0.12} & $20^3\times 64$ & 0.2 \\
$\approx${\bf 0.12} & $20^3\times 64$ & 0.4 \\
$\approx${\bf 0.12} & $24^3\times 64$ & 0.1 \\
$\approx${\bf 0.09} & $28^3\times 96$ & 0.2 \\
$\approx${\bf 0.09} & $32^3\times 96$ & 0.15 \\
\hline
\end{tabular}
\end{minipage}
\begin{minipage}{0.48\linewidth}
\centering
\begin{tabular}{lll}
\hline
$a$(fm) & size& $m_l/m_h$\\
\hline
$\approx${\bf 0.09} & $40^3\times 96$ & 0.1 \\
$\approx$0.09 & $64^3\times 96$ & 0.05 \\
$\approx$0.06 & $48^3\times 144$ & 0.2 \\
$\approx$0.06 & $48^3\times 144$ & 0.4 \\
$\approx$0.06 & $56^3\times 144$ & 0.14 \\
$\approx$0.06 & $64^3\times 144$ & 0.1 \\
$\approx$0.045 & $64^3\times 192$ & 0.2 \\
\hline
\end{tabular}
\end{minipage}
\caption{Summary of all ensembles to be included in the full
analysis. The ensembles in bold have been analyzed for this report.
Light(strange) sea quark masses are denoted by $m_l
(m_h)$.}
\end{table}

On each ensemble we compute the three-point correlation functions
relevant to the weak matrix elements and the two-point functions
relevant to the propagation of the $B$ and $D$ mesons.  For the
heavy-light mesons we use heavy clover quarks in the Fermilab
interpretation and light, improved staggered (asqtad) quarks
\cite{Bernard:1999xx,Bazavov:2009bb}. We use two types of
interpolating operators for the $B$ and $D$ mesons, namely local and
smeared using the 1S Richardson wave function.  We set the
valence-quark masses equal to the sea-quark masses. The bare lattice
charm and bottom quark masses are fixed by matching the kinetic
masses of the $D_s$ and $B_s$ mesons, respectively, to their
experimental values.  Some small adjustments will be needed to
refine this tuning but are not yet included in this preliminary
analysis.

In terms of interpolating operators $\mathcal{O}_B$ and $\mathcal{O}_D$
for the $B$ and $D$ mesons and the vector current, the two-point
and three-point functions are given by
\begin{eqnarray}
C^{2pt}(t;{\bf p})&=&\sum_{\bf x}\exp(i{\bf p}\cdot {\bf
x})\langle\mathcal {O}({\bf x},t)\mathcal {O}^{\dag}({\bf
0})\rangle\nonumber\\
&=&\sum_n s_n^t Z_n({\bf p}) [\exp(-E_n({\bf p})t) + \exp(-E_n({\bf
p})(N_t-t))]
 \\
 C_{V_\mu}^{3pt,B\rightarrow D}(t,T;
{\bf p})&=&\sum_{\bf x,\bf y}\exp(i{\bf p}\cdot {\bf
y})\langle\mathcal {O}_D(0)V_{\mu}({\bf y}, t)\mathcal
{O}_B^{\dag}({\bf x}, T)\rangle\ ,
\end{eqnarray}
where $s_n = \pm 1$ accounts for contributions that oscillate in
$t$, and $N_t$ is the lattice extent in the $t$ dimension.  Note
that in the three-point function above, we have put the $D$ meson at
the origin and the $B$ meson at $(\textbf{x}, T)$.

The matrix elements we want from Eq.~(\ref{threepts1}) to
Eq.~(\ref{threepts}) are obtained by factorization and reduction of
the three-point functions.  The reduction is based on the overlap
coefficients $Z_n({\bf p})$, energies and masses obtained in fits to
the two-point functions. For example, the $R_+({\bf p})$ matrix
element is obtained from
\begin{equation}
R_+(t,T;{\bf p})=\frac{C_{V_4}^{3pt,B\rightarrow D}(t,T;{\bf p})}
{\sqrt{C_{V_4}^{3pt,D\rightarrow D}(t,T;{\bf
0})C_{V_4}^{3pt,B\rightarrow B}(t,T;{\bf 0})}} \sqrt{\frac{Z_D({\bf
0})E_D}{Z_D({\bf
p})m_D}}e^{E_Dt-\frac{1}{2}m_DT}e^{-m_B(t-\frac{1}{2}T)}
\label{ratiofit}
\end{equation}
in the limit $t\rightarrow \infty$ and $T - t \rightarrow \infty$.  In
that limit it is a constant, $R_+({\bf p})$, independent of $t$.
However, at finite $t$ and $T$, complications from oscillating terms
and excited states must be taken into account.


Because they contain a light staggered quark, the meson
interpolating operators excite even- as well as odd-parity channels.
The unwanted even-parity states manifest themselves as terms that
oscillate in $t$, corresponding to terms in the two-point and
three-point functions with $s_n = -1$. To suppress the effect of the
oscillating terms in the three-point functions, we average in both
$t$ and $T$ (after removing the dominant exponential dependence on
$t$ as in Eq.~(\ref{ratiofit})) as follows:
\begin{eqnarray}
\overline{R}\equiv\frac{1}{2}R(0,t,T)
+\frac{1}{4}R(0,t,T+1)+\frac{1}{4}R(0,t+1,T+1)\ .
\end{eqnarray}
Here, $R$ refers to any of the quantities derived from three-point
functions, such as $R_+$, $R_-$, and $x_f$.

\begin{figure}
\centering
\includegraphics[width=0.4\textwidth]{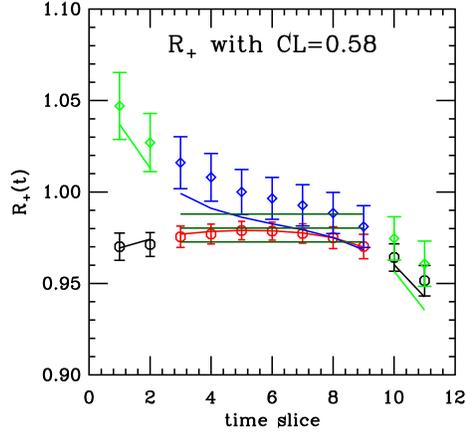}
\caption{ A sample fit to $R_+(t,T;\mathbf{p})$.  Example
three-point correlation fit for the 0.12 fm 0.14 $m_l/m_h$ ensemble.
The circles correspond to the local source, and diamonds, the
smeared source.  The fit result is obtained by a simultaneous fit to
both correlators using Eq.~(\protect\ref{ratiofit}). The horizontal
lines represent the central value and error of the result for
$R_+(\mathbf{p})$.} \label{fig:Reps}
\end{figure}

Having suppressed the contributions of the oscillating states to the
three-point correlators, we must still account for contributions from
excited states.  For example, for $D \rightarrow D$, the leading
contributions are
\begin{eqnarray}
&&C_{V4}^{3pt D\rightarrow
D}(p,t)=\sqrt{Z_D(p)}\frac{e^{-E_Dt}}{\sqrt{2E_D}} \langle
D|V_4|D\rangle\frac{e^{-m_D(T-t)}}{\sqrt{2m_D}}\sqrt{Z_D(0)}+\sqrt{Z_{D^{\prime}}(p)}\frac{e^{-E_{D^{\prime}}
t}}{\sqrt{2E_{D^{\prime}}}}\nonumber\\ &&\times\langle
D^{\prime}|V_4|D\rangle\frac{e^{-m_D(T-t)}}{\sqrt{2m_D}}\sqrt{Z_D(0)}+\sqrt{Z_D(p)}\frac{e^{-E_D
t}}{\sqrt{2E_D}}\langle
D|V_4|D^{\prime}\rangle\frac{e^{-m_{D^{\prime}}(T-t)}}{\sqrt{2m_{D^{\prime}}}}\sqrt{Z_{D^{\prime}}(0)}\
,
\end{eqnarray}
where the primes indicate the overlaps and energies of the excited
states of the same parity.  We have neglected the doubly excited
contribution, since the singly excited contribution is found already
to be small.  Thus, at finite $t$ and $T$ the ratio $R_+(t,T;{\bf
  p})$ in Eq.~(\ref{ratiofit}) is not constant, but has a small
contamination from excited states.  We take this into account by
fitting the resulting ratio to
\begin{equation}
  R_+({\bf p},t,T) = R_+({\bf p}) \exp(\delta m\, t) +
   A \exp(-\Delta E_D\, t) + B \exp(-\Delta m_B\,(T-t))
\end{equation}
as illustrated in Fig.~\ref{fig:Reps}.  The fit parameter $\delta m$
should be zero,  but it is introduced to allow for a small error in
determining $E_D$ and $m_B$. The other terms involve $\Delta E_D =
E_{D^\prime} - E_D$ and $\Delta m_B = m_{B^\prime} - m_B$.  These
parameters are obtained from the fits of two-point functions. The
results of two-point function fits then become prior central values
and widths for fitting $R_+(t,T;{\bf p})$. We use a similar method
to obtain $R_-({\bf p})$ and $x_f({\bf p})$.

\section{\label{sec:Chiral} Chiral and continuum extrapolation}

\begin{figure}
\centering
\includegraphics[width=0.45\textwidth]{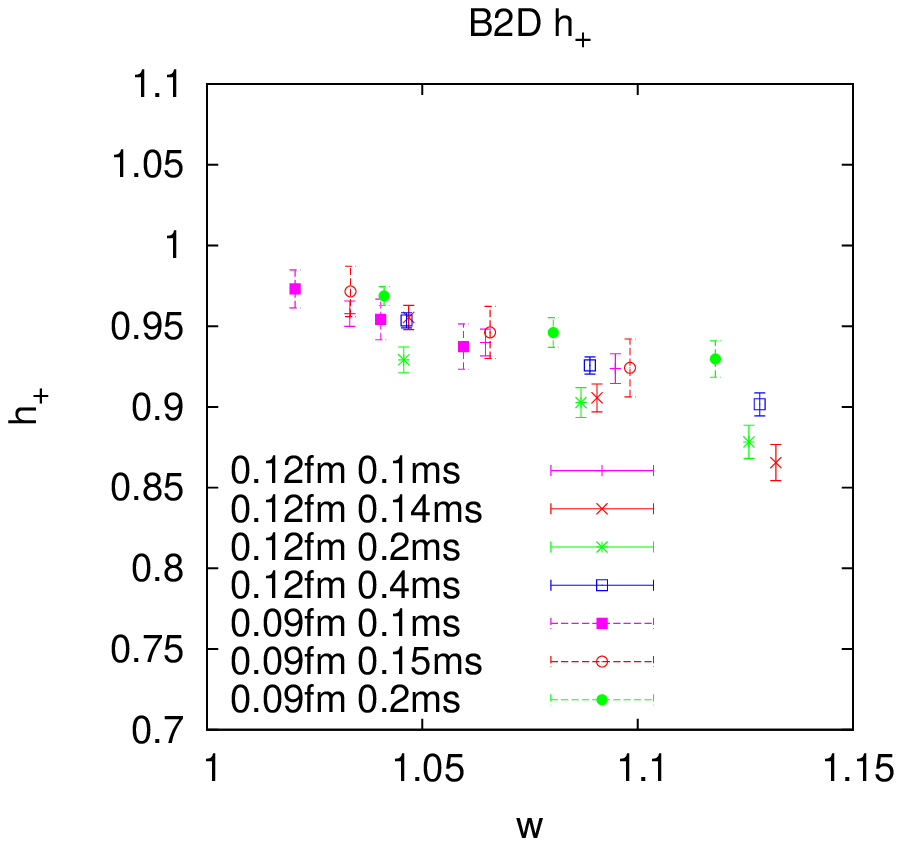}
\includegraphics[width=0.45\textwidth]{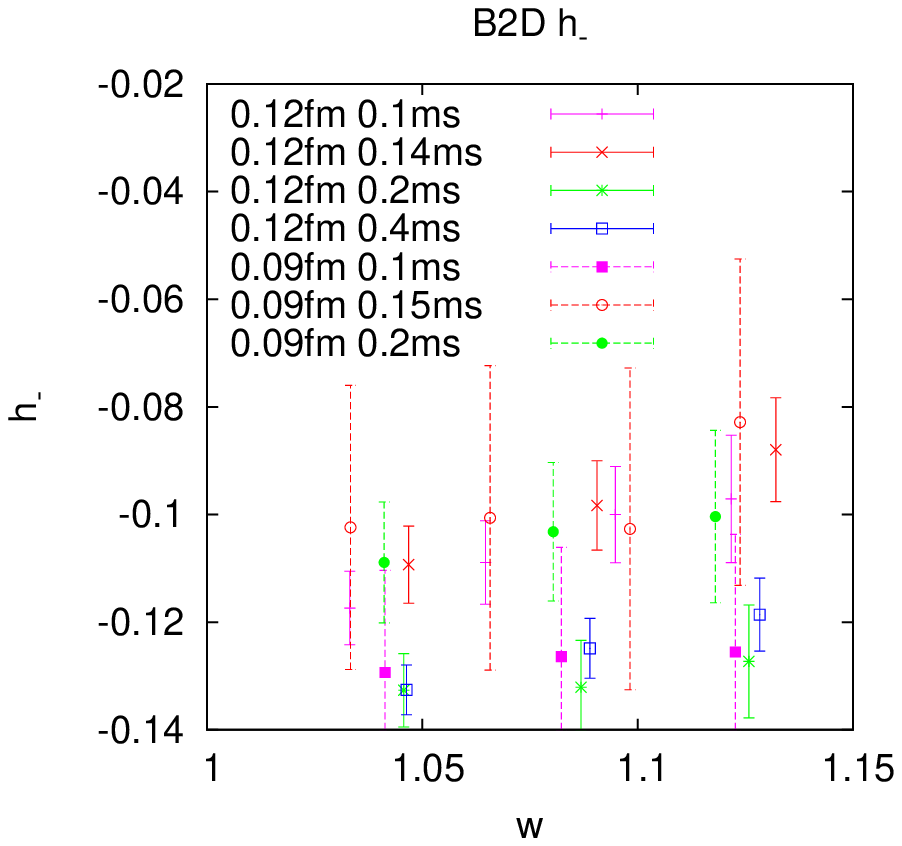}
\caption{\label{fig:fi3} Form factors $h_+$ (left) and $h_-$
(right) vs.\ $w$, omitting the matching factor $\rho_V$, on the bold
ensembles in Table~\protect\ref{table1}.  The color code is given in the legends.}
\end{figure}

To complete the analysis we will extrapolate in the light quark
masses to their physical values and take the continuum limit, using
staggered chiral perturbation theory \cite{Jack,Jack1} . At this
preliminary stage, we don't show physical values and chiral fit in
this report. The numerical data points for $h_+$ and $h_-$ are shown
in Fig.~\ref{fig:fi3}. We will need these data points to finish the
extrapolation of physical value of $h_+$ and $h_-$ and the chiral
fit.

\section{\label{sec:Fut} Future plans}

To complete the project, we need to (1) finish analyzing the
available ensembles, (2) make adjustments for the tuning of the
heavy quark masses, (3) compute the current renormalization
($\rho_V$ factor), and (4) combine the experimental results for the
differential decay rate with our results for the form factor to
obtain $|V_{cb}|$, using methods employed for $|V_{ub}|$ to extend
$w$ over the full kinematic range \cite{Ruth}.


%
%
\section*{Acknowledgments}
Computations for this work were carried
out with resources provided by the USQCD Collaboration, the Argonne
Leadership Computing Facility, the National Energy Research
Scientific Computing Center, and the Los Alamos National Laboratory,
which are funded by the Office of Science of the U.S. Department of
Energy; and with resources provided by the National Center for
Supercomputing Applications and the National Institute for
Computational Science, the Pittsburgh Supercomputer Center, the San
Diego Supercomputer Center, and the Texas Advanced Computing Center,
which are funded through the National Science Foundation's
Teragrid/XSEDE Program. This work was supported in part by the U.S.
Department of Energy under Grants No.~DE-FG02-91ER40664 (D.D.),
No.~DE-FG02-91ER40677 (D.D), and No.~DE-FC06-ER41446(C.D.); and in part by the U.S. National Science
Foundation under grants PHY0757333 (C.D.) and PHY0903571 (S.-W.Q.).
J.L. is supported by the STFC and by the Scottish Universities
Physics Alliance.
This manuscript has been co-authored by employees of Brookhaven Science
Associates, LLC, under Contract No. DE-AC02-98CH10886 with the
U.S. Department of Energy.
R.S.V. acknowledges support from BNL via the
Goldhaber Distinguished Fellowship.
D.D. was supported in part by
the URA Visiting Scholars' program at Fermilab. Fermilab is
operated by Fermi Research Alliance, LLC, under Contract No.
DE-AC02-07CH11359 with the United States Department of Energy.

\end{document}